\documentclass[twocolumn,flushbottom,showpacs,preprintnumbers,superscriptaddress,amsmath,amssymb]{revtex4}
\usepackage{graphicx}
\usepackage{subfigure}
\usepackage{dcolumn}
\usepackage{bm}
\usepackage{amssymb}
\usepackage{amssymb}
\usepackage{color}
\usepackage{subfigure}
\def\>{\rangle}

\begin{document}

\title{Metrology with $\mathcal{PT}$-symmetric cavities: Enhanced sensitivity near the $\mathcal{PT}$-phase transition}



\author{Zhong-Peng Liu}
\affiliation{Department of Automation, Tsinghua University,
Beijing 100084, P.~R.~China}
\affiliation{Tsinghua National Laboratory for Information Science
and Technology, Beijing 100084, P.~R.~China}
\author{Jing Zhang}\email{jing-zhang@mail.tsinghua.edu.cn}

\affiliation{Department of Automation, Tsinghua University,
Beijing 100084, P.~R.~China}
\affiliation{Tsinghua National Laboratory for Information Science
and Technology, Beijing 100084, P.~R.~China}
\affiliation{Department of Electrical
and Systems Engineering, Washington University, St.~Louis, MO
63130, USA}
\affiliation{CEMS, RIKEN, Saitama 351-0198, Japan}
\author{\c{S}ahin Kaya \"{O}zdemir}\email{ozdemir@ese.wustl.edu}
\affiliation{Department of Electrical and Systems Engineering,
Washington University, St.~Louis, MO 63130, USA}
\author{Bo Peng}
\affiliation{Department of Electrical and Systems Engineering,
Washington University, St.~Louis, MO 63130, USA}
\author{Hui Jing}
\affiliation{CEMS, RIKEN, Saitama 351-0198, Japan}
\affiliation{Department of Physics, Henan Normal University,
Xinxiang 453007, P.~R.~China}
\author{Xin-You L\"{u}}\affiliation{CEMS, RIKEN, Saitama 351-0198, Japan}
\author{Chun-Wen Li}
\affiliation{Department of Automation, Tsinghua University,
Beijing 100084, P.~R.~China}
\affiliation{Tsinghua National Laboratory for Information Science
and Technology, Beijing 100084, P.~R.~China}
\author{Lan Yang}\affiliation{Department of Electrical
and Systems Engineering, Washington University, St.~Louis, MO
63130, USA}
\author{Franco Nori}
\affiliation{CEMS, RIKEN, Saitama 351-0198, Japan}
\affiliation{Physics Department, The University of Michigan, Ann
Arbor, MI}
\author{Yu-xi Liu}
\affiliation{Tsinghua National Laboratory for Information Science
and Technology, Beijing 100084, P.~R.~China}
\affiliation{Institute
of Microelectronics, Tsinghua University, Beijing 100084, P.~R.~China}

\date{\today}

\begin{abstract}
We propose and analyze a new approach based on parity-time ($\mathcal{PT}$) symmetric microcavities with balanced gain and loss to enhance the performance of cavity-assisted metrology.
We identify the conditions under which $\mathcal{PT}$-symmetric microcavities allow to improve sensitivity beyond what is achievable in loss-only systems. We discuss its application to
the detection of mechanical motion, and show that the sensitivity is significantly enhanced in the vicinity of the transition point from unbroken- to broken-$\mathcal{PT}$ regimes. We believe that our results open a new direction for $\mathcal{PT}$-symmetric physical systems and it may find use in ultra-high precision metrology and sensing.
\end{abstract}


\maketitle

\emph{Introduction.---} The measurement of physical quantities
with high precision is the subject of metrology. This has
attracted much attention due to the increasing interest in, e.g.,
gravitational wave detection~\cite{RSchnabel}, sensing of
nanostructures~\cite{Zhao,Shi}, as well as global positioning and
navigation~\cite{ESBurillo,PMarks}. Developments in metrology over
the past two decades have provided the necessary tools to
determine the fundamental limits of measuring physical quantities
and the resources required to achieve
them~\cite{HMWiseman,GYXiang}.

Among many different approaches, cavity-assisted metrology (CAM),
where a high-quality ($Q$) factor cavity or resonator is coupled
to a device under test (DUT), has emerged as a versatile and
efficient experimental approach to achieve high-precision
measurements. In CAM, the coupling between the resonator and the
DUT manifests itself as a back-action-induced resonance frequency
shift, resonance mode splitting, or a sideband in the output
transmission spectrum~\cite{Clerk}. Cavity-assisted metrology has
been successfully applied for reading out the state of a
qubit~\cite{Siddiqi}, measuring tiny mechanical
motions~\cite{Aspelmeyer,Dobrindt,Guerlin,Szorkovszky,Galland,Nunnenkamp,Rabl,Liao},
and detecting nanoparticles with single-particle
resolution~\cite{JZhuNatPhonics,HeNatNanotechnology}.

\begin{figure}[]
\centerline{
\includegraphics[width=8 cm]{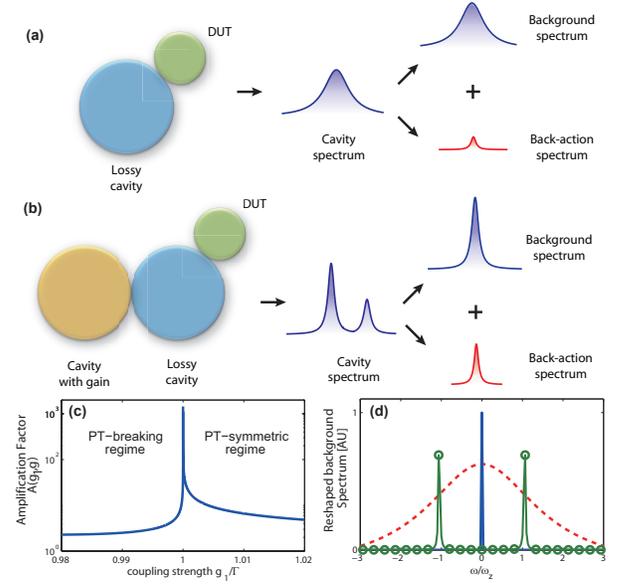}}
\caption{(Color online) (a) Schematic diagram of a single-cavity
transducer. Output spectrum is the sum of the cavity spectrum and
the back-action spectrum of the device under test (DUT). (b)
$\mathcal{PT}$-cavity-assisted transducer. The DUT is coupled to
two coupled cavities, one of which is active and the other is
passive. The cavity with gain introduced sharpens the background
Lorentz spectrum and amplifies the back-action
spectrum.(c) Amplification factor $A(g_1,g)$ of the
back-action spectrum versus the normalized coupling strength
$g_1/\Gamma$. Near the phase-transition point, i.e., $g_1=\Gamma$,
$A(g_1,g)$ increases very sharply, reaching values as high as
$1000$. (d) Normalized background spectra ${S}_a(\omega)$ of a
single cavity (red dashed curve) and the $\mathcal{PT}$-symmetric
cavities in the broken-$\mathcal{PT}$ regime (blue curve) and the
$\mathcal{PT}$-symmetric regime (green curve with circular marks).
The background spectrum is narrowed by the gain-loss balance of
$\mathcal{PT}$-symmetric structure, and, in the
$\mathcal{PT}$-symmetric regime, the background spectrum shows
splitting.}\label{ptscheme}
\end{figure}

The readout signal (i.e., the transmission spectrum) of CAM is
determined by the sum between the background spectrum of the
cavity and the back-action spectrum of the DUT. The background
spectrum is determined by the $Q$ of the cavity whereas the
back-action spectrum is determined by the strength of the
cavity-DUT coupling (also dependent on $Q$) and the quantity to be
measured. A broad background spectrum masks the back-action
spectrum and decreases signal-to-noise ratio (SNR)
[Fig.~\ref{ptscheme}(a)]. A higher coupling-strength between the
cavity and the DUT and a higher $Q$ of the cavity will be helpful
to detect very weak signals and enable to resolve fine structures
in the output spectra [Fig.~\ref{ptscheme}(b)]. A higher $Q$ is
also necessary to enhance the coupling strength between the cavity
and the DUT. For example, for optomechanical resonators, the
detection of tiny motions requires a strong optomechanical
coupling, which is only possible with an high $Q$-factor.
Therefore, CAM will benefit significantly from a narrower
background spectrum which is fundamentally limited by the material
absorption loss of the cavity. Techniques that will help to reach
the fundamental detection limit and measure very weak signals with
existing cavities are being actively sought.

We show that the performance of CAM is significantly enhanced if
the passive cavity (i.e., lossy; without optical gain) of the CAM
is coupled to an auxiliary cavity with optical gain (i.e., active
cavity) that balances the loss of the passive cavity. Such coupled
systems with balanced gain and loss are referred to as parity-time
($\mathcal{PT}$) symmetric systems~\cite{Bender}, which have been
widely studied both
theoretically~\cite{AASukhorukovPRA,HRamezaniPRA,ZLinPRL,XZhuOL,CHangPRL,GSAgarwal,HBenistyOE,NLazaridesPRL,YLumerPRL}
and
experimentally~\cite{AGuo,CERuterNatPhys,ARegensburgerNature,LFengNatMaterial,JSchindler,SBittner,CMBender,NBenderPRL,BPengNatPhys}.
As a specific application of $\mathcal{PT}$-CAM, we show the
enhancement in the detection of the motion of a nanomechanical
resonator in the proximity of the passive microcavity of the
$\mathcal{PT}$-CAM. The enhancement is significant in the vicinity
of the $\mathcal{PT}$-phase transition point through which the
system transits from broken- to unbroken-$\mathcal{PT}$ symmetry
and vice versa. The mechanism for the enhancement of the
measurement sensitivity in our system is attributed to two
features.
First, due to gain-loss balance we have almost lossless (extremely
high-$Q$) supermodes and hence much narrower background spectrum.
Thus, it becomes easier to resolve sideband induced by the DUT.
Second, effective interaction strength between the optical modes
and the DUT is significantly enhanced thereby allowing detection
of very weak perturbations in the DUT.

\emph{Cavity-assisted metrology (CAM) with
$\mathcal{PT}$-symmetric microcavities.---} The traditional CAM
system is composed of a lossy resonant optical cavity coupled to a
DUT [Fig.~\ref{ptscheme}(a)]. The interaction Hamiltonian is given
by $H_{\rm int}=ga^{\dag}a{z}$, where $a$ is the annihilation
operator of the cavity; ${z}$ is the DUT's observable being
measured; and $g$ is the coupling strength between the DUT and the
cavity. In order to realize the $\mathcal{PT}$-CAM, an active
cavity is directly coupled to the lossy cavity connected
[Fig.~\ref{ptscheme}(b)].
Here we consider the $\mathcal{PT}$-symmetric optical
microcavities discussed in Ref. \cite{BPengNatPhys}, where the
inter-cavity coupling strength is controlled by the distance
between the cavities, and the gain-to-loss ratio of the cavities
is adjusted by the power of the optical pump of the active cavity.
The Hamiltonian describing this $\mathcal{PT}$-CAM is
\begin{equation}\label{Hamiltonin for metrology PT system}
H=\left(\Delta-i\kappa\right)a^{\dag}a+\left(\Delta+i\gamma\right)c^{\dag}c+ga^{\dag}az+g_1(a^{\dag}c+c^{\dag}a),
\end{equation}
where $c$ is the annihilation operator of the active cavity;
$\kappa$ and $\gamma$, respectively, denote the loss and gain
rates of the passive and active cavities; $g_1$ is the
inter-cavity coupling strength; and $\Delta$ corresponds to the
effective frequencies of the two cavities.

Without the interaction term $ga^{\dag}az$, Eq.(\ref{Hamiltonin
for metrology PT system}) accounts for the coupling between the
optical modes of the microcavities and leads to two supermodes
$a_\pm$ that are described with the complex frequencies
$\omega_\pm=\Omega_\pm-i\Gamma_\pm=\Delta-i\chi\pm\beta$, where
$\beta=\sqrt{g^2_1-\Gamma^2}$, $\chi=(\kappa-\gamma)/2$, and
$\Gamma=(\gamma+\kappa)/2$. Clearly, $\beta$ experiences a
transition from a real to an imaginary value and vice versa at
$\Gamma=g_1$, where $\beta=0$. At the transition point,  the
complex eigenfrequencies coalesce, that is
$\omega_\pm=\Delta-i\chi$.

For $\Gamma<g_1$, $\beta$ is real (denoted as $\beta=\beta_r$) and
hence the complex eigenfrequencies of the supermodes become
$\omega_\pm=\Delta\pm\beta_r-i\chi$, implying that two supermodes
have different resonance frequencies [i.e.,
$\Omega_-\neq\Omega_+$] but the same damping rates and linewidths
[i.e., $\Gamma_\pm=\chi$]. The separation of resonance frequencies
is then given by $2\beta_r$. However, for $\Gamma>g_1$, $\beta$ is
imaginary, that is $\beta=i\beta_r$ and hence the complex
eigenfrequencies of the supermodes become
$\omega_\pm=\Delta-i(\chi\mp\beta_r)$, implying that two
supermodes are degenerate in their resonance frequencies [i.e.,
$\Omega_\pm=\Delta$] but have different damping rates and
linewidths quantified by $\Gamma_\pm=\chi\mp\beta_r$.

When the gain $\gamma$ of the active cavity balances the loss
$\kappa$ of the passive cavity, $\beta=0$ corresponds to a {\it
$\mathcal{PT}$-phase transition point}, where the supermodes
coalesce, and $\chi$ becomes zero, implying a lossless system. The
regime defined by $\Gamma<g_1$ corresponds to a {\it
$\mathcal{PT}$-symmetric phase}, where the lossless supermodes are
split by $2\beta_r$. The regime defined by $\Gamma>g_1$ denotes
the {\it broken $\mathcal{PT}$-phase} where the supermodes have
the same resonance frequencies, but one of them becomes lossy and
the other becomes amplifying.

Rewriting the non-Hermitian Hamiltonian (\ref{Hamiltonin for
metrology PT system}) in the supermode picture, we find that the
effective coupling strength between the supermodes and the DUT can
be written as~\cite{Supp}
\begin{equation}\label{Effective coupling}
g_{\rm eff}=\frac{g}{2\sqrt{|g_1^2-\Gamma^2|}}.
\end{equation}
Clearly, in the vicinity of the transition point $g_1=\Gamma$, the
effective coupling strength $g_{\rm eff}$ is significantly larger
than $g$. Therefore, the sensitivity is drastically enhanced near
the transition point.

The mechanism of the $\mathcal{PT}$-enhanced sensitivity can be
seen in a more quantitative manner by calculating the output
spectrum. Intuitively, the gain-loss balance of the
$\mathcal{PT}$-symmetric system makes the background spectrum
narrower, and the cavity with gain in the $\mathcal{PT}$ system,
which works as a dynamical amplifier, amplifies the back-action
spectrum of the DUT and thus increases the sensitivity of the
measurement.
In the regime of weak coupling between the cavity and the DUT, we
can omit the back-action of the cavity on the DUT. Then the
normalized spectrum can be written as~\cite{Supp}
\begin{equation}
{S(\omega)}\approx G(\omega)\left[{S}_a(\omega)+A(g_1,g){S}_z(\omega)\right]\label{mg}.
 \end{equation}
Here ${S}_a(\omega)$ is the single-cavity background Lorentz
spectrum calculated
by setting the inter-cavity coupling strength $g_1=0$.
Here ${S}_z(\omega)=\mathcal{F}\left[f(t)\mathcal{F}^{-1}S_{zo}(\omega)\right]$ is
the back-action spectrum from the DUT, with $S_{zo}(\omega)$
representing the spectrum of the DUT, and $\mathcal{F}$ and
$\mathcal{F}^{-1}$ are the Fourier and inverse Fourier transforms,
respectively. The time-domain function $f(t)$ is a form factor
that will broaden the back-action spectrum~\cite{Supp}.

Equation (\ref{mg}) implies that the $\mathcal{PT}$ structure and
the presence of the interaction between the lossy cavity and the
DUT give rise to an amplification factor $A(g_1,g)$ acting on the
back-action spectrum $S_z(\omega)$ [see Fig.~\ref{ptscheme}(c)].
In the $\mathcal{PT}$-breaking regime, when $g_1$ increases,
$A(g_1,g)$ first increases slowly from a very small value and
then, near the $\mathcal{PT}$ phase transition point, $A(g_1,g)$
increases very sharply, reaching a very high value. When $g_1$ is
further increased and the system enters the
$\mathcal{PT}$-symmetric regime, the amplification factor
$A(g_1,g)$ drops sharply to a small value and continues
decreasing, albeit with very slow rate as $g_1$ is increased.

In Fig.\ref{ptscheme}(d) we show the background spectrum
${S}_a(\omega)$ for a lossy cavity (red dashed curve) and for a
$\mathcal{PT}$ structure in the $\mathcal{PT}$-symmetric (green
curve with circles) and in the broken-$\mathcal{PT}$-symmetric
regimes (blue curve). Clearly, due to the presence of gain, the
susceptibility coefficient $G(\omega)$ reshapes ${S}_a(\omega)$,
leading to a background spectrum which is significantly narrower
than that of a single lossy cavity. Note that in the
$\mathcal{PT}$-symmetric regime, the background spectrum is split
into two due to the strong coupling between the cavities, and that
split resonance peaks seen in the spectrum is narrower than the
resonance of the single lossy cavity. Combining the narrower
background spectrum ${S}_a(\omega)$ in a $\mathcal{PT}$ structure
with the very high amplification factor in the vicinity of the
$\mathcal{PT}$-phase transition point will certainly lead to a
significantly enhanced sensitivity for the CAM.

\begin{figure}[]
\centerline{
\includegraphics[width=8 cm]{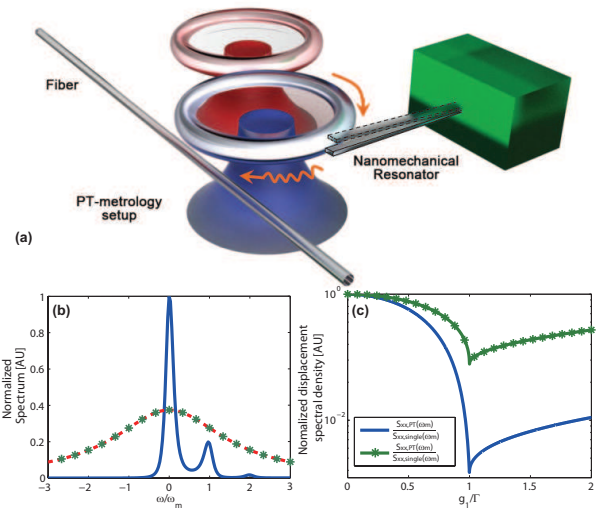}}
\caption{(Color online) (a)$\mathcal{PT}$-symmetric optomechanical
transducer realized by two coupled microtoriod resonators, i.e., a
silica microtoriod (blue; passive) and a $\rm Er^{3+}$-doped
silica microtoriod (red; active). The two-microtoriod setup is
coupled to a mechanical oscillator, e.g., a nanomechanical beam or
a cantilever, via the optical evanescent field.(b) The output
spectra of optomechanical transducers for $\omega_m/\kappa=0.3$:
single-cavity (red dashed curve), $\mathcal{PT}$-system in the
vicinity of the transition point (blue curve), and
two-lossy-cavity system in the vicinity of exceptional point
({EP}) (green curve with star marks). (c) The displacement
spectral densities $S_{\rm xx,PT}(\omega_m)$ and  $S_{\rm
xx,EP}(\omega_m)$ of the $\mathcal{PT}$-optomechanical transducer
(blue curve) and the two-lossy-cavity transducer (green curve with
star marks) normalized by the single-cavity displacement spectral
density $S_{\rm xx,single}(\omega_m)$ when $\omega_m/\kappa=0.3$.
The measurement sensitivity is enhanced for at least two order of
magnitude by the $\mathcal{PT}$ optomechanical transducer near the
transition point, i.e., $S_{\rm xx,PT}(\omega_m)/S_{\rm
xx,single}(\omega_m)<10^{-2}$, and the $\mathcal{PT}$-system
performs much better than the EP-system due to the gain-loss
balance of $\mathcal{PT}$ structure.
}\label{Es}
\end{figure}

\emph{Optomechanical transducer by $\mathcal{PT}$ breaking.---}
Here let us discuss an optomechanical transducer operated in the
vicinity of the $\mathcal{PT}$-phase transition point. We consider
that the lossy cavity of the $\mathcal{PT}$ structure supports a
mechanical mode. The Hamiltonian of this $\mathcal{PT}$
optomechanical system is obtained from Eq.~(\ref{Hamiltonin for
metrology PT system}) by replacing the operator $z$ by $b+b^\dag$,
where $b$ is the annihilation operator of the mechanical mode.
Similar to Eq.~(\ref{mg}), we can obtain the output spectrum of
this $\mathcal{PT}$ optomechanical system and compare it with the
single-cavity case [see Fig.~\ref{Es}(b)]


Here, let us consider a particular $\mathcal{PT}$-symmetric
optomechanical transducer realized by microtoriod resonators
showed in Fig.~\ref{Es}(a). This $\mathcal{PT}$-metrology system
consists of two coupled microtoriod-resonators, where one is a
silica microtoriod which acts as a passive cavity and the other is
a $\rm Er^{3+}$-doped silica microtoriod which is taken as the
active cavity~\cite{BPengNatPhys}. The microtoriod-based
$\mathcal{PT}$-metrology system is used to detect the tiny motion
of the mechanical oscillator, e.g., a nanomechanical beam or
cantilever, via the optical evanescent field of the passive
microtoroid~\cite{Anetsberger}.
In order to clearly show the differences between $\mathcal{PT}$
and a single-lossy-cavity optomechanics, we carried out numerical
simulations using experimentally accessible values of system
parameters:
$\Delta=0$, $\omega_m=6$ MHz, $\kappa=20$
MHz, $\gamma_m=0.2$ MHz, $\gamma=16$ MHz, $g_1=19.8$ MHz, and $g=5$
MHz. Since we have deliberately chosen the optomechanical coupling
strength $g$ very small, the susceptibility coefficient
of the single-cavity optomechanical system becomes very small, and
thus the back-action spectrum of the mechanical oscillator
is
masked by the background spectrum
of the cavity [see red dashed curve in Fig.~\ref{Es}(b)]. The
output spectrum of the $\mathcal{PT}$-symmetric transducer [blue
curve in Fig.~\ref{Es}(b)] shows two distinct features
originating from the gain-loss balance and the amplification
mechanism. First, the background spectrum is narrower and the
resonance peak located at $\omega/\omega_m=0$ has a higher value
than the single loss-cavity peak. Second, the back-action spectrum
of the mechanical motion is clearly seen as a sideband peak
sitting on the background spectrum at $\omega/\omega_m=1$. One can
also see the second-order mechanical sideband as a smaller peak
located $2\omega_m$ away from the main peak of the background
spectrum. These confirm that the $\mathcal{PT}$-symmetric
structure operated in the vicinity of the $\mathcal{PT}$-phase
transition point has the potential ability to detect very weak
mechanical motion.

To show the enhancement of the measurement-sensitivity by the
$\mathcal{PT}$-symmetric structure, we compare the displacement
spectral densities $S_{\rm xx,PT}(\omega)$ and $S_{\rm xx,single}(\omega)$
of the $\mathcal{PT}$-symmetric transducer
and the single-lossy-cavity transducer. In
fact, the displacement spectral density $S_{\rm xx,PT}(\omega)$
and the backaction force spectral density $S_{\rm FF,PT}(\omega)$ of the
$\mathcal{PT}$ optomechanical transducer can be calculated
as~\cite{Supp,Dobrindt}:
\begin{eqnarray}
&S_{\rm xx,PT}(\omega)=\frac{\Gamma_{-}^2\hbar\Omega_-}{64g_{\rm
eff}^2P_{\rm in}}\left(1+\frac{4\omega}{\Gamma_{-}^2}\right),&\label{sensitivity1}\\
&S_{\rm FF,PT}(\omega)=\frac{16\hbar g_{\rm eff}^2P_{\rm
in}}{\Gamma_-^2\Omega_-}\left(1+\frac{4\omega}{\Gamma_-^2}\right)^{-1},&\label{sensitivity2}
\end{eqnarray}
where $P_{\rm in}$ is the input power. It can be shown that
$S_{\rm xx,PT}(\omega)$ and $S_{\rm FF,PT}(\omega)$ satisfy the
Heisenberg inequality~\cite{Braginsky}: $\sqrt{S_{\rm
xx,PT}(\omega)S_{\rm FF,PT}(\omega)}\geq\hbar/2$, which means that
it is possible to obtain a smaller displacement spectral density
when we increase the backaction force spectral density. As shown in
Eq.~(\ref{sensitivity1}), the displacement spectral density
$S_{\rm xx,PT}(\omega)$ is proportional to the decay rate $\Gamma_-$ of the
supermode  and inversely proportional to the square of
the effective optomechanical coupling strength $g_{\rm eff}$.
Since $g_{\rm eff}$ can be efficiently increased in the vicinity
of the $\mathcal{PT}$-transition point [see Eq.~(\ref{Effective
coupling})] and the damping rate $\Gamma_-$ of the supermode  will
be decreased by the gain-loss balance of the
$\mathcal{PT}$-symmetric structure, thus the displacement spectral
density $S_{\rm xx,PT}(\omega)$ of the $\mathcal{PT}$
optomechanical transducer will beat those limits given by the
single-cavity case. This is confirmed by the numerical results
shown in Fig.~\ref{Es}. Here $S_{\rm xx,PT}(\omega)$ can be more than
two-order of magnitude smaller than the displacement spectral
density $S_{\rm xx,single}\left(\omega\right)$ of the single cavity.


Finally to show the effect of the presence of the gain in the
structure, we compare the sensitivity of a system, formed by two
coupled lossy cavities having loss rates of ${\kappa_1}$ and
$\kappa\geq\kappa_1$ with that of the $\mathcal{PT}$-symmetric
system formed by coupling a lossy cavity of loss rate $\kappa$
with a gain cavity of gain rate $\gamma$ \cite{Supp}. Note that
the loss rates of the cavities coupled to the mechanical mode is
the same (i.e., $\kappa$) for both systems. For the system formed
by two lossy cavities, there also exists a degenerate point where
the eigenfrequencies and the corresponding eigenstates of the
system coalesce. This transition point is generally known as the
exceptional point (EP) and has been studied in detail within the
field of non-Hermitian Hamiltonian~\cite{loss-2014}. For such a
system the EP takes place at $g_1=(\kappa-\kappa_1)/2$. In the
vicinity of an EP, we have $g_{\rm eff}\gg g$, as can be deduced
from Eq.~(\ref{Effective coupling}). For
$g_1>(\kappa-\kappa_1)/2$, the supermodes are split by
$2\sqrt{g_1^2-(\kappa-\kappa_1)^2/4}$ but have the same damping
rates $(\kappa+\kappa_1)/2$, whereas for
$g_1<(\kappa-\kappa_1)/2$, the supermodes are degenerate at
frequency $\Delta$ but have different damping rates
${(\kappa+\kappa_1)/2\mp\sqrt{(\kappa-\kappa_1)^2/4-g_1^2}}$.
Similar to the analysis for the $\mathcal{PT}$-system, we derived
the output spectrum in the vicinity of the EP and the
amplification factor~\cite{Supp}. We find the ratio of the
amplification factor of the $\mathcal{PT}$-system to the
two-lossy-cavity system to be
${4\gamma^2(\kappa+\kappa_1)/(\kappa-\gamma)^3}$, which implies
that the amplification factor of the $\mathcal{PT}$-system is
higher and the ratio approaches infinite if the gain-to-loss ratio
in the $\mathcal{PT}$-system is well-balanced. In
Fig.~\ref{Es}(b), we see that the normalized spectrum for the
two-lossy-cavity (curve with green stars) is similar to that of
the single-cavity system when the damping rates of the cavities
are the same, and thus the back-action spectrum of the mechanical
motion cannot be detected by the two-lossy-cavity. We also compare
the displacement spectral density of $\mathcal{PT}$-system with
that of EP-system in Fig.~\ref{Es}(c). We can observe similar decrease of the
displacement spectral density for $\mathcal{PT}$ and EP systems
near the transition point, but the $\mathcal{PT}$-system performs
much better because the effective damping rate of the
$\mathcal{PT}$-system is much smaller due to the gain-loss balance
of the $\mathcal{PT}$-symmetric structure.

\emph{Conclusion.---} We have proposed $\mathcal{PT}$-metrology as a efficient approach to improve the sensitivity and detection limit of cavity-assisted metrology beyond what is attainable in the conventional settings, where the system to-be-measured is coupled to a lossy cavity. In $\mathcal{PT}$-metrology a second cavity with gain is coupled to the lossy cavity to compensate its loss, thereby increasing the quality factor of the effective optical mode that is used to detect the weak signal, and enhancing the effective coupling strength between the cavity and the DUT. The enhancement is remarkable especially in the vicinity of the transition point where the supermodes of the system coalesce in their frequencies, and can be further enhanced as the gain-to-loss ratio approaches 1. We have showed that it is possible to realize an ultra-sensitive optomechanical transducer whose {\it sensitivity is at least two order of magnitude} better than the conventional single-cavity optomechanical transducers. We believe that this approach can be used for improving the performance of nanoparticle sensors \cite{Jan}, navigation systems, gravity-wave detectors and other cavity-assisted detection schemes.

YXL and JZ are supported by the National Basic Research Program of
China (973 Program) under Grant No. 2014CB921401, the Tsinghua
University Initiative Scientific Research Program, and the
Tsinghua National Laboratory for Information Science and
Technology (TNList) Cross-discipline Foundation. JZ is supported
by the NSFC under Grant Nos.~61174084, 61134008, 60904034. YXL is
supported by the NSFC under Grant Nos.~10975080, 61025022,
91321208. F.N. is partially supported by the RIKEN iTHES Project,
MURI Center for Dynamic Magneto-Optics via the AFOSR award number
FA9550-14-1-0040, and a Grant-in-Aid for Scientific Research (A).
LY is supported by ARO Grant No. W911NF-12-1-0026.

\end{document}